\documentclass[aps,prd,preprint,showpacs]{revtex4}
\usepackage[dvips]{graphicx}
\usepackage{bm}
\usepackage{amsmath}
\usepackage{amssymb}
\newcommand{\nn}{\nonumber}

\begin{document}

% Use the \preprint command to place your local institutional report
% number in the upper righthand corner of the title page in preprint mode.
% Multiple \preprint commands are allowed.
% Use the 'preprintnumbers' class option to override journal defaults
% to display numbers if necessary
\preprint{IC/HEP/04-1}

%Title of paper
\title{On the non-vanishing of the Collins mechanism \\ for single spin asymmetries}

\author{Elliot Leader}
\email{e.leader@imperial.ac.uk}

\affiliation{Imperial College London, Prince Consort Road, London
SW7 2BW, UK }

\date{\today}

\begin{abstract}
The Collins mechanism provides a non-perturbative explanation for
the large single-spin asymmetries found in hard semi-inclusive
reactions involving a transversely polarized nucleon. However,
there are seemingly convincing reasons to suspect that the
mechanism vanishes, and indeed it does vanish in the naive parton
model where a quark is regarded as an essentially ``free''
particle. We give an intuitive analysis which highlights the
difference between the naive picture and the realistic one and
shows how the Collins mechanism arises when the quark is described
as an off-shell particle by a field in interaction.

\end{abstract}

\pacs{13.60.Hb, 13.85.Ni, 13.87.Fh, 13.88.+e, 13.30.-a, 12.38.-t}
% insert suggested keywords - APS authors don't need to do this
%\keywords{}

\vspace{3cm}

Submitted to Physical Review D

%\maketitle must follow title, authors, abstract, \pacs, and \keywords
\maketitle

% body of paper here - Use proper section commands
% References should be done using the \cite, \ref, and \label commands
%\section{}
% Put \label in argument of \section for cross-referencing
\section{\label{1}Introduction}

The Collins mechanism ~\cite{1} provides a possible explanation of
the very large single-spin asymmetries -- sometimes as big as 40\%
-- found in semi-inclusive reactions like $ p^\uparrow
p\rightarrow \pi X$ or $e p^\uparrow \rightarrow e\pi X$, where
$p^\uparrow$ is a transversely polarized proton. The existence of
these asymmetries has been known for decades and one of the great
puzzles has been the fact that it is quite impossible to produce
such large asymmetries via perturbative QCD mechanisms. The
Collins mechanism is a  ``soft'' non-perturbative effect
describing the fragmentation of a transversely polarized quark
into e.g. $\pi + X$. It makes the angular distribution of the pion
with respect to the direction of motion of the quark depend upon
the direction of the quark's transverse polarization. But this
feature is itself puzzling, since it is well known ~\cite{2,3}
that in the decay of a transversely polarized particle or
resonance $ R^\uparrow \rightarrow \pi + X$ the angular
distribution $ W(\theta ,\phi )$ of the pion cannot depend on the
direction of the transverse polarization of $ R $ , in any parity
conserving theory. Indeed, as we shall see explicitly, the Collins
mechanism vanishes in the naive parton model where the quark is
visualized as a ``free'' particle. The Collins effect thus depends
crucially on treating the quark as off-shell and described by an
interacting field.\\
It was at one time thought that the time-reversal invariance of
QCD could be utilized to prove the vanishing of the Collins
mechanism, but Collins  ~\cite{1} pointed out a flaw in the
argument based upon a failure to distinguish between so-called
\textit{in} and \textit{out} states in field theory. Although this
mathematical argument shows that the Collins mechanism need not
vanish, it does not provide an intuitive picture of what is
happening.\\
In this paper we give an intuitive analysis of the Collins
mechanism based upon rotational invariance, which makes clear the
difference between particle or resonance decay, which
\textit{cannot} have a Collins effect, and quark fragmentation,
which can.\\
In Section 2 we revisit the standard QCD picture for hard
semi-inclusive reactions, stressing the difference between
treating the active quark as ``free'' and as an off-shell
interacting particle. Section 3 derives the angular distribution,
contrasting the cases of ``free'' and interacting quarks. The
results are discussed in Section 4. Some technical details are
relegated to an Appendix.

\section{QCD analysis of ${\bf e \, p^\uparrow \to e \, \pi \, X}$}

\begin{figure}[h]

 \begin{center}
 \includegraphics[width=12cm,height=6cm]{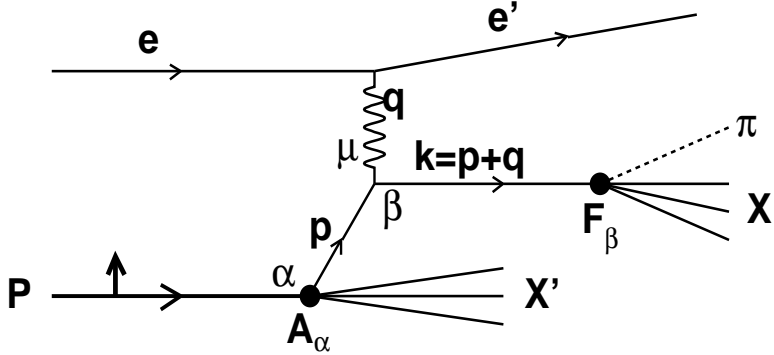}
 \end{center}
 \caption{Feynman diagram for semi-inclusive scattering\label{a}}
 \end{figure}

For concreteness we shall present our analysis in the context of
semi-inclusive deep inelastic lepton-proton scattering ( see
Fig.~\ref{a} ) in which an unpolarized lepton interacts with a
transversely polarized proton and the final state pion is detected
at angles $ (\theta , \phi )$ in the``$ \gamma $" p CM frame, with
the proton moving along the OZ axis.
%%%%%%%%%%%%%%%%%%%%%%%%%%%%%%%%%%%%%%%%%%%%%%%%%%%55
%put figure picture here
%Here is an example of the general form of a figure:
% Fill in the caption in the braces of the \caption{} command. Put the label
% that you will use with \ref{} command in the braces of the \label{} command.
% Use the figure* environment if the figure should span across the
% entire page. There is no need to do explicit centering.

%%%%%%%%%%%%%%%%%%%%%%%%%%%%%%%%%%%%%%%%%%%%%%%%%%%%%%%%

Regarding Fig.~\ref{a} as a Feynman diagram, the amplitude for the
hadronic part of the process is given by

\begin{equation}\label{2.1}
  H^{\mu}= F_{\beta} ( \gamma^{\mu}) _{\beta\alpha }A_{\alpha}
\end{equation}

where $A_{\alpha}$ is the amplitude for the proton, momentum $P$,
to emit a quark with four-momentum $p$ and Dirac index $\alpha$,
and $F_{\beta}$ is the amplitude for a quark of four-momentum $k$
and Dirac index $\beta$ to fragment into a pion and some other
particle $X$. As is usual in the parton model, both $A_{\alpha}$
and $F_{\beta}$ are defined to include their adjacent quark
propagators.\\
In QCD these amplitudes are given by:

\begin{equation}\label{2.2}
  A_{\alpha}= \int d^{4}x e^{ip.x} <X'\mid \Psi_{\alpha}(x)\mid
  P>
  \end{equation}

  \begin{equation}\label{2.3}
  F_{\beta}=\int d^{4}z e^{-ik.z} <\pi ,X\mid
  \bar\Psi_{\beta}(z)\mid 0>
\end{equation}

where $k_{\mu}$ is a time-like four-vector, $k^{0}>0$, and the
hadronic tensor $W^{\mu\nu}$ is then

\begin{equation}\label{2.4}
  W^{\mu\nu} = \sum_{X,X'} H^{\mu}H^{\nu *}
\end{equation}

Note that the $\pi X$ final state is, strictly speaking, an
\textit{out} state, but this is irrelevant in our analysis.\\
In these and all further expressions we are only interested in the
general structure of the amplitudes, so we will ignore numerical
and other inessential factors Also, since we deal with unpolarized
leptons, we are only concerned with the part of $W^{\mu\nu}$
symmetric under $\mu \leftrightarrow \nu$.

Substituting Eq.~(\ref{2.1}) into Eq.~(\ref{2.4}) we obtain

\begin{equation}\label{2.5}
 W^{\mu\nu} = \rho^{\mu\nu}_{\alpha\beta} \sum_{X} F_{\alpha}
 \bar{F}_{\beta}
\end{equation}

where, as usual,

\begin{equation}\label{2.6}
  \bar {F}_{\beta} = F^{*}_{\tau}\gamma^{0}_{\tau\beta}
\end{equation}

and

\begin{equation}\label{2.7}
  \rho^{\mu\nu}_{\alpha\beta} = (\gamma^{\mu}\Phi
  \gamma^{\nu})_{\alpha\beta}
\end{equation}

with $\Phi$ the usual quark-quark correlation function

\begin{equation}\label{2.8}
  \Phi_{\alpha\beta} = \int d^{4}x e^{-ip.x}<P\mid
  \bar{\Psi}_{\beta}(x)\Psi_{\alpha}(0)\mid P>
\end{equation}

Since the labels $\mu,\nu$ play no role in the following, aside
from the symmetry under $\mu \leftrightarrow \nu$, we shall
suppress them in the following, simply writing the LHS of
Eq.~(\ref{2.7}) as $\rho_{\alpha\beta}$.

Consider now the structure of the $ q\rightarrow \pi X$
fragmentation amplitude

\begin{eqnarray}
 F_{\beta} &\equiv & F_{\beta}(p_{\pi};X,\lambda_{X};k)
\nonumber \\
            &=& \int d^{4}z
e^{-ik.z}<p_{\pi};X,\lambda_{X}\mid
             \bar{\Psi}_{\beta}(z)\mid 0> \label{2.9}
\end{eqnarray}
where $\lambda_{X}$ is the helicity of $X$.\\
If $\bar{\Psi}(z)$ were a free field $\bar{\Psi}^{0}(z)$ we could
make the usual Fourier expansion into creation and annihilation
operators

\begin{eqnarray}\label{2.10}
  \bar{\Psi}_{\beta}^{0}(z) & \sim &
  \sum_{\lambda}\int\frac{d^{3}p}{E_{p}}\{\bar{u}_{\beta}(
  \bm{p},\lambda )a^{\dag}_{0}( \bm{p},\lambda )e^{ip.z} \nn \\
  & + & \bar{v}_{\beta}( \bm{p},\lambda )b_{0}( \bm{p},\lambda
  )e^{-ip.z}\}
\end{eqnarray}

where $E_{p}=\sqrt{\bm{p}^{2}+m^{2}}$, and the subscript $0$ on
the operators signifies free field operators.\\
When $\bar{\Psi}(z)$ is a field in interaction we can still make a
3-dimensional Fourier expansion at some instant of time, say
$t=0$. Then one finds ~\cite{4}, at arbitrary times $t$,

\begin{eqnarray}
\bar{\Psi}_{\beta}(z)& \sim & \sum_{\lambda}\int
\frac{d^{3}p}{E_{p}}\{
\bar{u}_{\beta}(\bm{p},\lambda)a^{\dag}(\bm{p},\lambda
;t)e^{-i\bm{p}.\bm{z}}
\nonumber \\
& + &
\bar{v}_{\beta}(\bm{p},\lambda)b(\bm{p},\lambda;t)e^{i\bm{p}.\bm{z}}
\} \label{2.11}
\end{eqnarray}

where now the time dependence is non-trivial, being controlled by
the Hamiltonian $H$:

\begin{eqnarray}
a^{\dag}(\bm{p},\lambda;t)=e^{iHt}a^{\dag}
(\bm{p},\lambda)e^{-iHt}
\nonumber \\
b(\bm{p},\lambda;t)=e^{iHt}b(\bm{p},\lambda)e^{-iHt} \label{2.12}
\end{eqnarray}

This is to be contrasted with the free field case where

\begin{eqnarray}
a^{\dag}_{0}(\bm{p},\lambda;t)=e^{iE_{p}t}a^{\dag}_{0}(\bm{p},\lambda)
\nonumber \\
b_{0}(\bm{p},\lambda;t)=e^{-iE_{p}t}b_{0}(\bm{p},\lambda)
\label{2.13}
\end{eqnarray}

Of course the interacting $a^{\dag}(\bm{p},\lambda;t)$ and $
b(\bm{p},\lambda;t)$ are not simply one-particle creation and
annihilation operators. However, the spinors occurring in
Eqs.~(\ref{2.10}) and ~(\ref{2.11}) are the same as in the free
case i.e. are mass
m spinors.\\
Bearing in mind Eq.~(\ref{2.11}) we have for the field operator
appearing in Eqs.~(\ref{2.3}) and ~(\ref{2.9})

\begin{eqnarray}
\bar\Psi_{\beta}(k)\equiv \int
d^{4}ze^{-ik.z}\bar{\Psi}_{\beta}(z) & = &
\frac{1}{E_{k}}\sum_{\lambda} \{
\bar{u}_{\beta}(\bm{k},\lambda)A^{\dag}(k,\lambda)\nonumber \\
& + & \bar{v}_{\beta}(-\bm{k},\lambda)B(k,\lambda) \} \label{2.14}
\end{eqnarray}

where

\begin{eqnarray}
A^{\dag}(k,\lambda) \equiv \int dt
e^{-ik_0t}a^{\dag}(\bm{k},\lambda;t) \nn \\
B(k,\lambda)\equiv \int dt e^{-ik_0t}b(-\bm{k},\lambda;t)
\label{2.15}
\end{eqnarray}

As stressed earlier, these are not single particle creation and
annihilation operators. However, acting on the vacuum state $\mid
0>$ they produce states which, as is shown in the Appendix, are
eigenstates of momentum and energy with definite helicity and
definite parity. Thus we shall define

\begin{eqnarray}
\mid k^{\mu},\lambda)_{+} \equiv A^{\dag}(k,\lambda)\mid 0>
\label{2.16}
\end{eqnarray}

and

\begin{eqnarray}
\mid k^{\mu},\lambda)_{-}\equiv B(k,\lambda)\mid 0> \label{2.17}
\end{eqnarray}

and the key point will be the fact that the states in
Eqs.~(\ref{2.16}) and (\ref{2.17}) have \textit{opposite} parity,
as indicated by the labels +/-. We use the notation $\mid \quad )$
to emphasize the fact that these states are not necessarily one
particle states. Hence we may write Eq.~(\ref{2.9}) in the form

\begin{eqnarray}
F_{\beta}(p_{\pi};X,\lambda_{X};k)=
\frac{1}{E_{k}}\sum_{\lambda}\{ \bar
u_{\beta}(\bm{k},\lambda)<p_{\pi};X,\lambda_{X}\mid
k^{\mu},\lambda)_{+}  \nn \\
\bar v_{\beta}(-\bm{k},\lambda)<p_{\pi};X,\lambda_{X}\mid
k^{\mu},\lambda)_{-} \} \label{2.18}
\end{eqnarray}

This is to be contrasted with the expression one obtains for the
decay of a \textit{particle} of momentum $\bm k$ (or for the case
of a free quark field) namely

\begin{eqnarray}
F_{\beta}^{0}(p_{\pi};X,\lambda_{X};k)=\frac{1}{E_{k}}\sum_{\lambda}\bar
u_{\beta}(\bm k,\lambda)<p_{\pi};X,\lambda_{X}\mid k,\lambda>
\label{2.19}
\end{eqnarray}
The analogue of the second term on the RHS of Eq.(\ref{2.18}) is
here absent, because, via Eq.~(\ref{2.10}), it involves

\begin{eqnarray}
\int dt e^{-ik_{0}t}e^{-iE_{p}t}\propto \delta (k_{0}+E_{p})=0
\label{2.20}
\end{eqnarray}

since $k_{0}>0$.

\section{The hadron angular distribution}

It will suffice to consider the decay
$$ ``q" \longrightarrow  \pi + X $$
in the CM system of the $\pi$ and $X$ i.e. where $\bm p_{\pi}+\bm
p_{X}=0$, since using the properties of the states $\mid
k,\lambda)_{+/-}$ derived in the Appendix one can then derive the
angular distribution in any other frame.

Let us look first at the decay of a spin1/2 particle
(equivalently, the case of a ``free" quark) where Eq.~(\ref{2.19})
holds. From Eq.~(\ref{2.5}) and (\ref{2.19})

\begin{eqnarray}
W^{\mu\nu}\propto \rho^{uu}_{\lambda
'\lambda}\sum_{X}<p_{\pi};X,\lambda_{X}\mid
k,\lambda><p_{\pi};X,\lambda_{X}\mid k,\lambda '>^* \label{3.1}
\end{eqnarray}

where

\begin{eqnarray}
\rho^{uu}_{\lambda '\lambda}=\bar u_{\alpha}(k,\lambda
')\rho_{\alpha\beta}u_{\beta}(k,\lambda) \label{3.2}
\end{eqnarray}

Rotational and parity invariance allow us to write, for the matrix
element in Eq.~(\ref{3.1}), in the frame where $\bm p_{\pi}=-\bm
p_{X}$

\begin{equation}\label{3.3}
<p_{\pi};X,\lambda_{X}\mid k,\lambda>=M(\lambda_{X})e^{i\lambda
\phi}d^{1/2}_{\lambda \lambda_{X}}(\theta)
\end{equation}

where $(\theta,\phi)$ are the polar angles of $\bm p_{\pi}$ and
the reduced matrix element $M(\lambda_{X})$ is independent of
$\lambda$ ~\cite{3}, and

\begin{equation}\label{3.4}
M(-\lambda_{X})=\eta M(\lambda_{X}) \qquad (\eta =\pm 1)
\end{equation}

Then Eq.~(\ref{3.1}) becomes

\begin{equation}\label{3.5}
W^{\mu\nu}\propto \rho^{uu}_{\lambda '\lambda}e^{i\phi(\lambda
'-\lambda)}\sum_{\lambda_{X}}\mid M(\lambda_{X})\mid^2
d^{1/2}_{\lambda '\lambda_{X}}(\theta)d^{1/2}_{\lambda
\lambda_X}(\theta)
\end{equation}

For a spin1/2 particle moving along $OZ$ the dependence on the
transverse polarization is contained in the off diagonal helicity
matrix elements

\begin{equation}\label{3.6}
\rho^{uu}_{+-}=\rho^{{uu}^*}_{-+}=\frac{1}{2}(\textit{P}_{x}-i\textit{P}_{y})
\end{equation}

The contribution from $\rho^{uu}_{+-}$ to Eq.~(\ref{3.5}) is
\begin{equation}\label{3.7}
\rho^{uu}_{+-}e^{i\phi}\{\mid M(1/2)\mid^2d^{1/2}_{1/2
,1/2}(\theta)d^{1/2}_{-1/2, 1/2}(\theta)+\mid
M(-1/2)\mid^2d^{1/2}_{1/2, -1/2}(\theta)d^{1/2}_{-1/2,
-1/2}(\theta) \}
\end{equation}

\begin{equation}\label{3.8}
=\rho^{uu}_{+-}e^{i\phi}\mid
M(1/2)\mid^2\{d^{1/2}_{1/2,1/2}(\theta)d^{1/2}_{-1/2,1/2}(\theta)+d^{1/2}_{-1/2,-1/2}(\theta)d^{1/2}_{1/2,-1/2}(\theta)\}
\end{equation}

where we have used Eq.~(\ref{3.4}).

Now
\begin{equation}\label{3.9}
d^{J}_{-1/2,-1/2}(\theta)=d^{J}_{1/2,1/2}(\theta), \qquad
d^{J}_{1/2,-1/2}(\theta)=-d^{J}_{-1/2.1/2}(\theta)
\end{equation}

so that the expression in parenthesis in Eq.~(\ref{3.8}) vanishes.
Similarly the term arising from $\rho^{uu}_{-+}$ vanishes.

Hence, in the  decay of a particle or a ``free" quark
$``q"\rightarrow \pi + X$, the angular distribution of the pion is
independent of the transverse polarization of the decaying
particle, i.e. the Collins mechanism vanishes.

The crucial point leading to the above conclusion was our ability
to factor out the reduced matrix element terms $\mid
M(1/2)\mid^2=\mid M(-1/2)\mid^2$ in going from Eq.~(\ref{3.7}) to
Eq.~(\ref{3.8}). It is this step that fails in the case of an
interacting quark field. For the the analogoue of Eq.~(\ref{3.1})
is, on putting Eq.~(\ref{2.18}) into Eq.~(\ref{2.5}),

\begin{equation}\label{3.10}
W^{\mu\nu} \propto
\rho^{uu}_{\lambda'\lambda}\sum_{\lambda_{X}}\mathcal{M}_{+}\mathcal{M}^{\prime *}_{+}
+ \rho^{vv}_{\lambda'\lambda}\sum_{\lambda_{X}}\mathcal{M}_{-}\mathcal{M}^{\prime *}_{-}
 +  \rho^{uv}_{\lambda'\lambda}\sum_{\lambda_{X}}
\mathcal{M}_{-}\mathcal{M}^{\prime *}_{+}+
\rho^{vu}_{\lambda'\lambda}\mathcal{M}_{+}\mathcal{M}^{\prime *}_{-}
\end{equation}

where, for brevity we have used

\begin{eqnarray}\label{3.11}
\mathcal{M}_{+/-}\equiv <p_{\pi};X,\lambda_{X}\mid k,\lambda )
_{+/-} \nn \\
\mathcal{M}'_{+/-}\equiv <p_{\pi};X,\lambda_{X}\mid
k,\lambda')_{+/-}
\end{eqnarray}

and where, analogously to Eq.~(\ref{3.2}),

\begin{eqnarray}\label{3.12}
\rho^{vv}_{\lambda'\lambda}=\bar
v_{\alpha}(k,\lambda')\rho_{\alpha\beta}v_{\beta}(k,\lambda) \nn \\
\rho^{uv}_{\lambda'\lambda}=\bar
u_{\alpha}(k,\lambda')\rho_{\alpha\beta}v_{\beta}(k,\lambda) \nn
\\
\rho^{vu}_{\lambda'\lambda}=\bar
v_{\alpha}(k,\lambda')\rho_{\alpha\beta}u_{\beta}(k,\lambda)
\end{eqnarray}

The analogues of Eqs.~(\ref{3.3}) and (\ref{3.4}) are now

\begin{equation}\label{3.13}
\mathcal{M}_{+/-}=<p_{\pi};X,\lambda_{X}\mid
k,\lambda)_{+/-}=M_{+/-}(\lambda_{X})e^{i\lambda\phi}d^{1/2}_{\lambda,\lambda_{X}}(\theta)
\end{equation}

with

\begin{equation}\label{3.14}
M_{+}(-\lambda_{X})=\eta M_{+}(\lambda_{X}) \qquad
M_{-}(-\lambda_{X})=-\eta M_{-}(\lambda_{X})
\end{equation}

Consider now the dependence of the angular distribution upon the
transverse polarization of the quark i.e. upon the terms in
Eq.~(\ref{3.10}) involving $\rho_{+-}$ and $\rho_{-+}$. For the
terms containing $\rho^{uu}_{+-},\rho^{uu}_{-+},\rho^{vv}_{+-}$
and $\rho^{vv}_{-+}$ the argument is identical to the free
particle case and the Collins mechanism vanishes. But, for
example, for the $\rho^{uv}_{+-}$ term one has

\begin{eqnarray}
\rho^{uv}_{+-}e^{i\phi}\sum_{\lambda_{X}}M^*_{+}(\lambda_{X})M_{-}(\lambda_{X})d^{1/2}_{-1/2,\lambda_{X}}(\theta)
d^{1/2}_{1/2,\lambda_{X}}(\theta) = \nn \\
\rho^{uv}_{+-}e^{i\phi}\{M^*_{+}(1/2)M_{-}(1/2)d^{1/2}_{-1/2,1/2}(\theta)d^{1/2}_{1/2,1/2}(\theta)+
\nn
\\
M^*_{+}(-1/2)M_{-}(-1/2)d^{1/2}_{-1/2,-1/2}(\theta)d^{1/2}_{1/2,-1/2}(\theta)\}
\label{3.15}
\\
=\rho^{uv}_{+-}e^{i\phi}M^*_{+}(1/2)M_{-}(1/2)\{d^{1/2}_{-1/2,1/2}(\theta)d^{1/2}_{1/2,1/2}(\theta)
- d^{1/2}_{-1/2,-1/2}(\theta)d^{1/2}_{1/2,-1/2}(\theta)
\label{3.16}
\end{eqnarray}

where we have used Eq.~(\ref{3.14}).

Note the crucial difference: the minus sign in the parenthesis in
Eq.~(\ref{3.16}) compared with (\ref{3.8}). Consequently the
angular factor in Eq.~(\ref{3.16})
does \textit{not} vanish and
the Collins mechanism survives. The other contributions, from
$\rho^{uv}_{-+},\rho^{vu}_{+-}$ and $\rho^{vu}_{-+}$ also do not
vanish, and there is no cancellation between them.

Hence for the interacting, off-shell quark the Collins mechanism
does not vanish, and the fundamental reason is that the operator
in Eq.~(\ref{2.14}) creates a superposition of states with
opposite parity, whereas in the case of a free field it creates a
state with just one definite parity.

\section{Discussion}

We have, for simplicity, discussed the case of the semi-inclusive
production of pions, but, in fact, the results hold equally well
for any hadron $h$. Thus, in the fragmentation of a transversely
polarized off-shell quark $``q"\rightarrow h + X $ where the quark
is described by an interacting QCD field, the angular distribution
of the $h$ \textit{can} depend on the direction of the quark
transverse polarization i.e. the Collins mechanism is non-zero.

On the contrary, for a decaying particle, or when the quark is
regarded as a ``free" on-shell particle, the Collins mechanism
vanishes.

The key difference between the two cases is that the operator
involved in the fragmentation amplitude

\begin{equation}\label{4.1}
\int d^{4}ze^{-ik.z}<h;X,\lambda_{X}\mid \bar \Psi_{\beta}(z)\mid
0> \qquad  (k_{0}>0)
\end{equation}

i.e. the operator

\begin{equation}\label{4.2}
\bar \Psi_{\beta}(k)=\int d^{4}ze^{-ik.z}\bar \Psi_{\beta}(z)
\end{equation}

when acting on the vacuum creates a superposition of states with
opposite parity when it is an interacting field,

\begin{equation}\label{4.3}
\bar \Psi_{\beta}(k)\mid 0> =\frac{1}{E_{k}}\sum_{\lambda}\{ \bar
u_{\beta}(\bm{k},\lambda)\mid k,\lambda)_{+} + \bar
v_{\beta}(-\bm{k},\lambda)\mid k,\lambda)_{-} \}
\end{equation}

as can be seen from Eqs.~(\ref{2.14}) to ~(\ref{2.17}), whereas it
creates a state with definite parity when it is a free field. The
existence of a definite parity for the decaying particle is
crucial for the cancellation that causes the Collins mechanism to
vanish. Some attempts have been made to estimate the size of the
Collins effect via dynamical calculations. For access to this
literature, see ~\cite{5}.

\begin{figure}[!ht]

\begin{center}
\includegraphics[width=12cm,height=6cm]{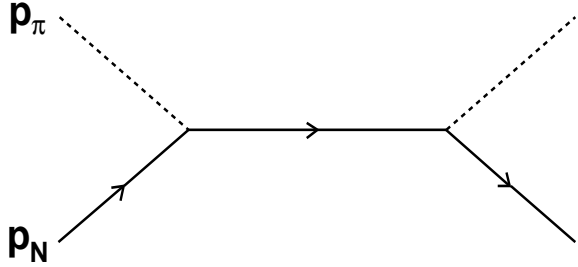}
\end{center}
\caption{Feynman diagram for $\pi N$ scattering \label{b}}
\end{figure}

It is interesting to note that there is a well known example of
this parity distinction between free on-shell and interacting
off-shell particles in the ancient literature ~\cite{6}. In $\pi
N$ elastic scattering the Born term shown in Fig.~(\ref{b}), is
known to contribute to both the $s$ and $p$-wave $\pi N$ states
i.e.to states of opposite parity. Only when

\begin{equation}\label{4.4}
s=(p_{\pi}+p_{N})^2\longrightarrow M^2
\end{equation}

where $M$ is the nucleon mass i.e. when the exchanged nucleon is
on the mass-shell, corresponding to the exchanged object being a
real particle with positive parity, does the $p$-wave amplitude
totally dominate.

Thus there is, in the end, no surprise that the non-vanishing of
the Collins mechanism is sensitive to whether the decaying quark
is treated as a ``free" on-shell particle or an interacting
off-shell one.

\begin{acknowledgments}
The author is grateful to the Royal Society of Edinburgh for an
Auber Bequest Fellowship, to Elena Boglione for help in preparing
the manuscript, and to Mauro Anselmino for reading it.
\end{acknowledgments}

\appendix*
\section{Properties of the states $\mid \lowercase{k},\lambda )_{+/-}$}

Let $ \hat{P}^{\mu}$ be the energy and momentum operators of the
theory. Then

\begin{equation}\label{A1}
[ \hat{P}^{\mu} , \bar \Psi (z) ] = -i\frac{\partial \bar
\Psi}{\partial z_{\mu}}
\end{equation}

Substituting Eq.~(\ref{2.11}) and then integrating with $\int
d^{4}ze^{-ik.z}$, we obtain for the momentum operators $ \hat{\bm
P}$

\begin{eqnarray}\label{A2}
\sum_{\lambda}\Big [\hat{\bm P}, \{ \bar u(\bm
k,\lambda)A^{\dag}(\bm
k,\lambda) + \bar v(-\bm k,\lambda)B(\bm k,\lambda) \}\Big ] = \nn \\
\sum_{\lambda}\{\bm k\bar u(\bm k,\lambda)A^{\dag}(\bm k,\lambda)
+ \bm k\bar v(-\bm k,\lambda)B(\bm k,\lambda)\}
\end{eqnarray}

Multiplying respectively by $\gamma^{0}u(\bm k,\lambda)$ and
$\gamma^{0}v(-\bm k,\lambda)$ to separate the $A^{\dag}$ and $B$
terms we get

\begin{equation}\label{A3}
[ \hat{\bm P}, A^{\dag}(k,\lambda)]  =  \bm k  A^{\dag}(
k,\lambda)
\end{equation}

\begin{equation}\label{A4}
[ \hat{\bm{P}}, B(k,\lambda)]  =  \bm k B(k,\lambda)
\end{equation}

from which follows that the states

\begin{eqnarray}\label{A5}
\mid k,\lambda)_{+} & = & A^{\dag}(k,\lambda)\mid 0>  \nn  \\
\mid k,\lambda )_{-} & = & B(k,\lambda)\mid 0>
\end{eqnarray}

are eigenstates of momentum, with momentum $\bm k$.

To study the energy we use the fact that

\begin{equation}\label{A6}
[ \hat{P_{0}}, a^{\dag}(\bm k,t)] = -i\frac{\partial a^{\dag}(\bm
k,t) }{\partial t}
\end{equation}

so that, from Eq.~(\ref{2.15})

\begin{equation}\label{A7}
[ \hat{P_{0}}, A^{\dag}(k,\lambda) ] = \int dte^{-ik_{0}t}\Big
(-i\frac{\partial a^{\dag}(\bm k,t)}{\partial t}\Big )
\end{equation}

Given that all such operator equations should really be considered
within the framework of test functions, we may integrate the RHS
of Eq.~(\ref{A7}) by parts, discarding the terms at $t=\pm
\infty$. Hence ~(\ref{A7}) becomes

\begin{equation}\label{A8}
[ \hat{P_{0}}, A^{\dag}(k,\lambda) ] = k_{0}A^{\dag}(k,\lambda)
\end{equation}

A similar relation holds for $B(k,\lambda)$. It follows that

\begin{equation}\label{A9}
\hat{P_{0}}\mid k,\lambda)_{+/-} = k_{0}\mid k,\lambda)_{+/-}
\end{equation}

Consider now the operation of space inversion

\begin{equation}\label{A10}
\mathcal P \bar \Psi (\bm z, t){\mathcal P}^{-1} = \bar \Psi (-\bm
z,t)\gamma_{0}
\end{equation}

It is well known that for free fields this leads to

\begin{eqnarray}\label{A11}
\mathcal P a^{\dag}_{0}(\bm k,\lambda){\mathcal P}^{-1} & = &
a^{\dag}_{0}(-\bm k,\lambda)  \nn  \\
\mathcal P b_{0}(\bm k,\lambda){\mathcal P}^{-1}& = & -b_{0}(-\bm
k,\lambda)
\end{eqnarray}

On comparing Eqs.~(\ref{2.10}) and (\ref{2.11}) it is clear, for
interacting fields, that the same results will hold for
$a^{\dag}(\bm k,\lambda;t)$ and $b(\bm k,\lambda;t)$, hence that

\begin{eqnarray}\label{A12}
\mathcal P A^{\dag}(k,\lambda){\mathcal P}^{-1} & = &
A^{\dag}(\tilde{k},\lambda)  \nn  \\
\mathcal P B(k,\lambda){\mathcal P}^{-1} & = &
-B(\tilde{k},\lambda)
\end{eqnarray}

where $\tilde{k}=(k_{0},-\bm k)$.

It follows that the states $\mid k,\lambda)_{+/-}$ have opposite
parity.

Finally, under a Lorentz transformation $l$

\begin{equation}\label{A13}
U(l)\bar \Psi_{\beta}(z)U(l^{-1}) = \bar
\Psi_{\alpha}(lz)D_{\alpha\beta}(l)
\end{equation}

where $D_{\alpha\beta}$ corresponds to the spinor transformation
matrix $S_{\alpha\beta}$ given in Bjorken and Drell~\cite{7}. If
we define, as in Eq.~(\ref{2.14}),

\begin{equation}\label{A14}
\bar \Psi_{\beta}(k) = \int d^{4}ze^{-ik.z}\bar \Psi_{\beta}(z)
\end{equation}

then

\begin{eqnarray}
U(l)\bar \Psi_{\beta}(k)U(l^{-1})& = & \int
d^{4}ze^{-ik.z}U(l)\bar
\Psi_{\beta}(z)U(l^{-1}) \nn  \\
& = & D_{\alpha\beta}(l)\int d^{4}ze^{-ik.z}\bar \Psi_{\alpha}(lz)
\nn  \\
& = & D_{\alpha\beta}(l)\int d^{4}z'e^{-ik.l^{-1}z'}\bar
\Psi_{\alpha}(z') \qquad (z'=lz)  \nn  \\
& = & D_{\alpha\beta}(l)\int d^{4}z'e^{-ilk.z'}\bar
\Psi_{\alpha}(z')  \nn  \\
& = & \bar \Psi_{\alpha}(lk)D_{\alpha\beta}(l)  \label{A15}
\end{eqnarray}

Using the expression Eq.~(\ref{2.14}) for $\bar \Psi_{\beta}(k)$
yields

\begin{eqnarray}
\bar \Psi_{\alpha}(lk)D_{\alpha\beta}(l) =
\frac{1}{E_{k'}}\sum_{\lambda} \{
\bar{u}_{\alpha}(\bm{k'},\lambda)A^{\dag}(lk,\lambda)\nonumber \\
+ \bar{v}_{\alpha}(-\bm{k'},\lambda)B(lk,\lambda)
\}D_{\alpha\beta}(l) \label{A16}
\end{eqnarray}

where

\begin{equation}\label{A17}
k'=lk
\end{equation}

Now~\cite{7}

\begin{equation}\label{A18}
\bar{u}_{\alpha}(\bm{k'},\lambda)D_{\alpha\beta}(l) =
\bar{u}_{\beta}(\bm{k},\lambda')\mathcal D^{1/2}_{\lambda
\lambda'}(r^{-1})
\end{equation}

and

\begin{equation}\label{A19}
\bar{v}_{\alpha}(\bm{k'},\lambda)D_{\alpha\beta}(l)
=\bar{v}_{\beta}(\bm{k},\lambda')\mathcal D^{1/2}_{\lambda'
\lambda}(r)
\end{equation}

where $r$ is the Wick helicity rotation (analogous to the Wigner
rotation). Thus

\begin{eqnarray}\label{A20}
\bar \Psi_{\alpha}(lk)D_{\alpha\beta}(l) =
\frac{1}{E_{k'}}\sum_{\lambda}\{\bar{u}_{\beta}(\bm{k},\lambda')\mathcal
D^{1/2}_{\lambda \lambda'}(r^{-1})A^{\dag}(lk,\lambda) \nn  \\+
\bar{v}_{\beta}(-\bm{k},\lambda')\mathcal D^{1/2}_{\lambda'
\lambda}(r)B(lk,\lambda)\}
\end{eqnarray}

Multiplying Eq.~(\ref{A13}) first by $[\gamma^{0}u(\bm
k,\lambda)]_{\beta}$ then by $[\gamma^{0}v(-\bm
k,\lambda)]_{\beta}$, and using (\ref{A14}) and (\ref{A20}) we
finally obtain

\begin{equation}\label{A21}
U(l)A^{\dag}(k,\lambda)U(l^{-1}) = \frac{E_{k}}{E_{k'}}\mathcal
D^{1/2}_{\lambda' \lambda}(r^{-1})A^{\dag}(lk,\lambda')
\end{equation}

\begin{equation}\label{A22}
U(l)B(k,\lambda)U(l^{-1}) = \frac{E_{k}}{E_{k'}}\mathcal
D^{1/2}_{\lambda \lambda'}(r)B(lk,\lambda')
\end{equation}

These imply that the states $\mid k,\lambda)_{+/-}$ transform just
like genuine particle states under rotations, and under boosts
differ from particle states only by the factor $
\frac{E_{k}}{E_{k'}}$, which is innocuous for our
purposes~\cite{8}.

Thus the states $\mid k,\lambda)_{+/-}$ possess all the properties
required in the derivation of the results in Sections II and III.

\bibliography{basename of .bib file}

\end{document}